\documentclass[review]{elsarticle}

\usepackage{color}
\usepackage{lineno,hyperref}
\usepackage{amssymb}
\usepackage{amsmath}
\usepackage{xspace}

\newcommand{\Jpsi}{\ensuremath{{\rm J}/\psi}\xspace}
\newcommand{\de}[1]{{\rm d}#1}

\begin{document}

\begin{frontmatter}

\title{
Solutions to the Balitsky-Kovchegov equation including the dipole orientation
}

\author[CVUT]{J. Cepila}
\author[CVUT]{J. G. Contreras}
\author[CVUT]{M. Vaculciak}
\address[CVUT]{Faculty of Nuclear Sciences and Physical Engineering, Czech Technical University in Prague, Czech Republic}

\begin{abstract}
Solutions of the target-rapidity Balitsky-Kovchegov (BK) equation are studied considering, for the first time, the complete impact-parameter dependence, including the orientation of the dipole with respect to the impact-parameter vector. 
In our previous work \cite{Cepila:2018faq} it has been demonstrated that the spurious Coulomb tails could be tamed using the collinearly-improved kernel and an appropriate initial condition in the projectile-rapidity BK equation. 
Introducing a different interpretation of the evolution variable, the target-rapidity formulation of the BK equation brings non-locality in rapidity and a kernel modification, removing the term that previously helped to suppress the Coulomb tails.
To address this newly emerged non-locality, three different prescriptions are explored here to take into account the rapidities preceding the initial condition. Two of these approaches induce mild Coulomb tails, while the other is free from this effect within the studied rapidity range. The range is chosen to correspond to that of interest for existing and future experiments. To demonstrate that this set up can be used for phenomenological studies, the obtained solutions are used to compute the $F_2$ structure function of the proton and the diffractive photo- and electro-production of J$/\psi$ off protons. The predictions agree well with HERA data, confirming that the target-rapidity Balitsky-Kovchegov equation with the full impact-parameter dependence is a viable tool to study the small Bjorken-$x$ limit of perturbative QCD at current facilities like RHIC and LHC as well as in future colliders like the EIC. 

\end{abstract}


\end{frontmatter}

\section{Introduction
\label{sec:intro}}
Deep-inelastic scattering (DIS) experiments allow for the study of the inner structure of hadrons within perturbative quantum chromodynamics (pQCD).
The  DIS cross sections measured by the H1 and ZEUS Collaborations at HERA~\cite{Abramowicz:2015mha}  indicate that the gluon distribution rises rapidly for gluons carrying a  small fraction $x$ of the proton momentum. The growth of the gluon distribution is driven by splitting processes which increase the number of gluons in the proton. When the gluon occupancy becomes large enough, recombination processes activate~\cite{Gribov:1983ivg,Mueller:1985wy} until a dynamical equilibrium, called gluon saturation, is reached. For a recent review see e.g. Ref.~\cite{Morreale:2021pnn}, or see Ref.\cite{Kovchegov:2012mbw} for an in-depth treatment of QCD at small $x$.

One of the tools to describe the evolution of the proton structure at high energies within pQCD is the Balitsky-Kovchegov (BK) equation~\cite{Balitsky:1995ub,Kovchegov:1999yj,Kovchegov:1999ua}, which describes the evolution in rapidity of the interaction between a colour dipole and a hadronic target. The original BK equation uses the projectile rapidity as the evolution variable. Recently, it was proposed to use the target rapidity ($\eta$)  as the evolution variable~\cite{Ducloue:2019ezk} in order to improve the stability of the equation by ensuring the correct time ordering of gluon emissions.  This formulation has the added advantage of having a direct relation with $x$, namely $\eta=\ln(x_0/x)$, where $x_0$ is the rapidity at which the BK evolution starts. The price to pay for these improvements is the introduction of non-local terms in the equation.

The interaction between the colour dipole and the hadronic target, that is the solution of the BK equation, is embodied by the dipole amplitude $N(\vec{r},\vec{b};\eta)$ which depends on two two-dimensional vectors defined in the plane transverse to the direction of dipole motion. They are $\vec{r}$, whose magnitude $r$ corresponds to the dipole size, and $\vec{b}$, the impact parameter between the dipole and the target. Most of the work done to date on finding solutions to the BK equation has been under the assumption of a dependence on $r$ only, which corresponds to the case of a large and homogeneous target. In this approximation, 
solutions of the BK equation with target rapidity evolution, $N(r;\eta)$, provide a good description of DIS at HERA~\cite{Ducloue:2019jmy,Beuf:2020dxl}. 

Up to now, the inclusion of the impact-parameter dependence in the solutions of the BK equation has been reported  only for the case of projectile evolution. The first studies in this direction~\cite{Golec-Biernat:2003naj} found a nonphysical behaviour of the dipole amplitude at large impact parameters, where the amplitude did not fall off as fast as expected from unitarity considerations. This phenomenon is known as {\em Coulomb tails} and it is attributed to confinement effects. This behaviour put a serious question mark on the feasibility of phenomenological applications. Later on, it was shown that with some ad-hoc modifications  to account for confinement, HERA data could be described~\cite{Berger:2010sh,Berger:2011ew,Berger:2012wx}. 
Soon thereafter, the kernel of the BK equation evolved in projectile rapidity was improved by the inclusion of corrections to take into account the resummation of gluon emission~\cite{Beuf:2014uia,Iancu:2015vea,Iancu:2015joa}. Using these developments as well as an appropriate initial condition, 
it was demonstrated, that the BK equation can be solved including the $r$ and $b\equiv|\vec{b}|$ dependence without the appearance of Coulomb tails, at least for energies relevant for current, and soon to come~\cite{Accardi:2012qut}, experimental results on deep-inelastic scattering and diffractive vector meson production off proton and nuclear targets~\cite{Cepila:2018faq, Bendova:2019psy,Cepila:2020xol,Bendova:2020hbb}. A similar behaviour  was also found in analytical solutions at large impact parameters~\cite{Contreras:2019vox}. Other approaches studied the impact-parameter dependence of the BK equation under the assumption of an SO(3) symmetry, see e.g. Ref.~\cite{Hagiwara:2016kam}. This corresponds to using a kernel similar to those explored in the first studies mentioned above~\cite{Golec-Biernat:2003naj}.

The suppression of Coulomb tails, which allowed for phenomenological applications of the impact-parameter BK equation, was due mainly to two factors: the collinearly improved kernel and choosing an appropriate initial condition. Rewriting the BK equation in target rapidities implies modifications to the kernel, in particular to the part that was responsible for the suppression of the Coulomb tails. This raises the question if the new kernel, and the introduced non-locality, are also able to suppress this unwanted effect; that is, if the solutions are viable for phenomenological applications.

Furthermore, for inhomogeneous targets it is expected that the dipole amplitude depends also on the angle, $\theta$, between $\vec{r}$ and $\vec{b}$, because the two ends of the dipole probe a different density of colour charges in the target. This effect has been studied in different frameworks, e.g. Refs.~\cite{Kopeliovich:2007fv,Hatta:2017cte,Iancu:2017fzn,Mantysaari:2020lhf,Dumitru:2021tvw,Kopeliovich:2021dgx}, but not yet with the target-rapidity BK equation.

In this Letter, we present numerical solutions to the target-rapidity Balitsky-Kovchegov equation including the impact-parameter dependence as well as the dependence on the angle between the dipole size and the impact-parameter vectors.  We find that the treatment of the non-local term for rapidities earlier than the point where the BK evolution starts has an influence on the behaviour of the Coulomb tails and offers a way to tame them, in which case a reasonable description of HERA data is achieved. This letter is organised as follows: an overview of the formalism is presented in Sec.~\ref{sec:overview}, the dipole amplitudes found as solutions of the BK equation are discussed in Sec.~\ref{sec:results}, while Sec.~\ref{sec:pheno}
shows the application of these solutions to compute physical observables and compares the obtained predictions with existing HERA data. 

\section{Brief overview of the formalism
\label{sec:overview}}

The BK equation in target rapidity~\cite{Ducloue:2019ezk} is 
\begin{align}
\frac{{\rm d}N(\vec{r}, \vec{b}; \eta)}{{\rm d}\eta} = \int \mathrm{d} \vec{r}_1 K(r, r_1, r_2) \Big[ 	&   N(\vec{r}_1, \vec{b}_1; \eta_1) + N(\vec{r}_2, \vec{b}_2; \eta_2) - N(\vec{r}, \vec{b}; \eta)	\nonumber \\
		& - N(\vec{r}_1, \vec{b}_1; \eta_1) N(\vec{r}_2, \vec{b}_2;\eta_2) \Big]. \label{eq:BK}
\end{align}
The first three terms on the right-hand-side of the equation take into account the splitting of a dipole at $(\vec{r},\vec{b})$ into two dipoles at $(\vec{r}_1,\vec{b}_1)$ and $(\vec{r}_2,\vec{b}_2)$, while  the last term represents the recombination of  two  dipoles. The
 rapidities $\eta_1$ and $\eta_2$ introduce the non-locality mentioned above; they are defined as 
 
 \begin{equation}
 \eta_j = \eta - \max\lbrace 0, \ln(r^2/r^2_j)\rbrace.
 \label{eq:shift}
 \end{equation}
 The  kernel is given by 
\begin{align}
    K(\Vec{r}, \Vec{r}_1, \Vec{r}_2) = \frac{\bar{\alpha}_s}{2\pi} \frac{r^2}{r_1^2r_2^2} \bigg[ \frac{r^2}{\min \lbrace r_1^2, r_2^2 \rbrace} \bigg]^{\pm \bar{\alpha}_s A_1}. \label{eq:kernel}
\end{align}
 The constant $A_1 = \frac{11}{12}$ and $J_1$ is the Bessel function. Here,
 $\bar{\alpha}_s = \frac{N_{\rm C}}{\pi} \alpha_s$ with \mbox{$\alpha_s = \alpha_s(\min\lbrace r, r_1, r_2 \rbrace)$} being the running strong coupling constant. It is evaluated, as in our previous works~\cite{Cepila:2018faq, Bendova:2019psy}, in the variable-number-of-flavours scheme. It depends on an infrared regulator, $C^2$. The number of colours is $N_{\rm C} = 3$. 

Note that a comparison of the kernel in Eq.~\ref{eq:kernel} and the collinearly improved kernel used in the projectile-rapidity approach, see Eq.~(9) in Ref.~\cite{Iancu:2015vea}, shows that the latter has an extra term accounting for the resummation of the double collinear logs to all orders. It was shown in our previous work, see Fig. 5 in Ref.~\cite{Bendova:2019psy}, that precisely this term is crucial to suppress the Coulomb tails. This observation motivates the question whether the new formulation of the BK equation produces viable solutions for phenomenological applications.

The initial condition for the start of the evolution is given by
\begin{equation}
N(\vec{r}, \vec{b}; \eta_0)  = 1 - \exp{\left(- \frac{1}{4}(Q_{s0}^2\,r^2)^\gamma \,T(b,r)\left\{1 + c\cos(2\theta)\right\}\right)},
\label{eq:N0}
\end{equation}
with 
\begin{equation}
T(r, b)= \exp{\left(-\frac{b^2+(r/2)^2}{2B}\right)}.
\label{eq:Trb}
\end{equation}
The parameter $Q_{s0}^2$ is related to the scale where saturation effects set in, $T(r, b)$ represents the target profile, the parameter B is related to the size of the hadronic target and $\gamma$ is the so-called anomalous dimension (see Ref.~\cite{Albacete:2004gw} for an early discussion in the context of the BK equation). The parameter $c$ controls the amount of the expected asymmetry on the $\theta$ dependence.

Three different approaches are considered to take into account the cases where the non-locality  shifts, see Eq.~(\ref{eq:shift}), produce a rapidity earlier than the initial rapidity $\eta_0 = \ln(x_0/x_0) = 0$:
\begin{itemize}
\item[A:] Suppress completely the contribution from rapidities earlier than $\eta_0$, namely, $N(\vec{r}, \vec{b}; \eta<\eta_0) = 0$. 
\item[B:] Smoothly suppress for a range of rapidity, specifically between $\eta_1=\ln(x_0/1)$ and $\eta_0$ the amplitude according to
\begin{equation}
N(\vec{r}, \vec{b}; \eta<\eta_0)  = 1 - \exp{\left(- \frac{1}{4}\left[(x_0/x)^{\lambda}\,Q_{s0}^2\,r^2\right]^\gamma \,T(b,r)\left\{1 + c\cos(2\theta)\right\}\right)},
\label{eq:N0min}
\end{equation}
and then $N(\vec{r}, \vec{b}; \eta<\eta_1) = 0$. The form of the suppression is inspired by the well-known GBW model~\cite{Golec-Biernat:1998zce}. 
\item[C:] Use the initial condition at all early rapidities: $N(\vec{r}, \vec{b}; \eta<\eta_0) = N(\vec{r}, \vec{b}; \eta_0) $. 
\end{itemize}

The BK equation is solved numerically in a fine logarithmic grid in $r$ and $b$ and a linear grid in $\theta$ using the Runge--Kutta method with the integrals performed with Simpson's method. The step in rapidity is 0.1. The parameter values that we have used in the numerical simulations shown below are listed in Table~\ref{tab:para}. 

\begin{table}[t!]
\caption{
\label{tab:para} 
Values of the parameters of the initial condition and of the strong coupling constant used for the numerical simulations. \\}
\centering
\begin{tabular}{ccccccc}
\hline
$x_0$ & $Q_{s0}^2$ & B & $\gamma$ & $\lambda$ & $c$ & $C$\\
0.01 & 0.496 GeV$^2$ & 3.8 GeV$^{-2}$ & 1.25  & 0.288 &1 & 30 \\
\hline
\end{tabular}
\end{table}

\section{Results
\label{sec:results}}

\begin{figure}[t!]
\centering 
\includegraphics[width=0.48\textwidth]{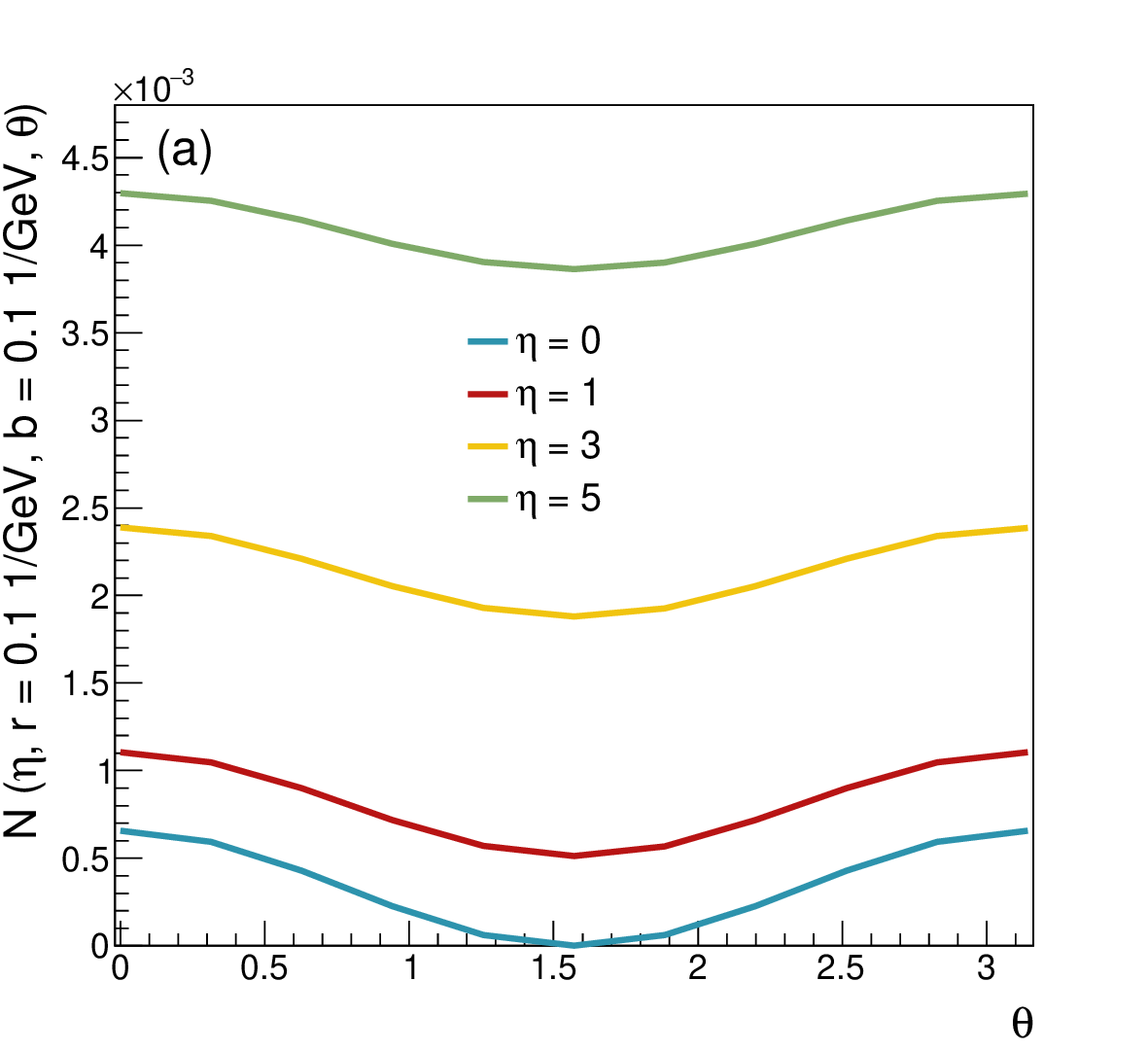}
\includegraphics[width=0.48\textwidth]{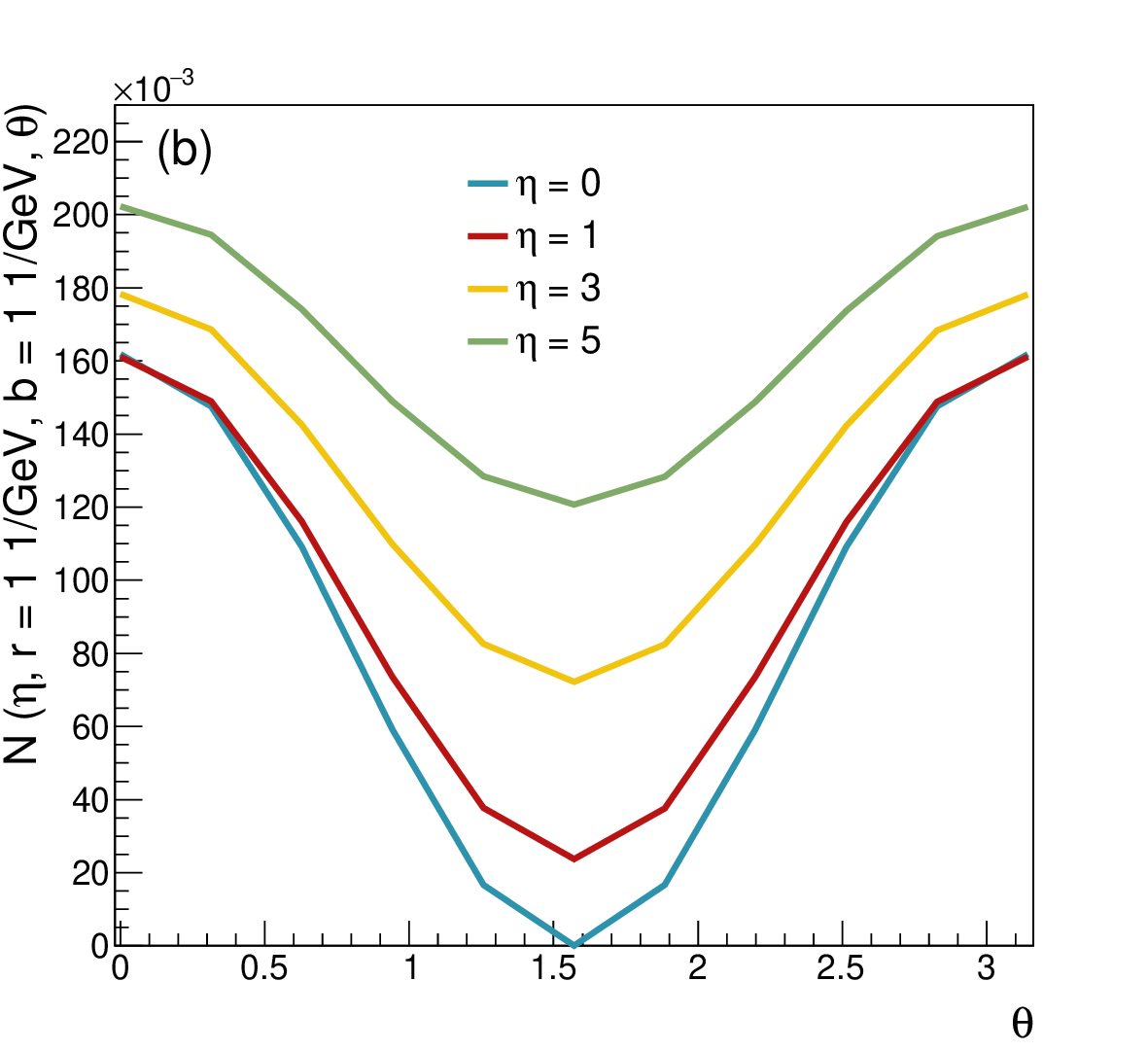}
\caption{Dependence of the dipole amplitude on $\theta$, the angle between the dipole-size and the impact-parameter vectors, (a) for a dipole of size $r=0.1$ GeV$^{-1}$ at impact parameter $b=0.1$~GeV$^{-1}$, and (b) a dipole of size $r=1$~GeV$^{-1}$ at impact parameter $b=1$~GeV$^{-1}$. Solutions are shown at different rapidities for the approach A to non-locality.
\label{fig:theta} 
}
\end{figure}

The dependence on $\theta$, the angle between the dipole-size and the impact-parameter  vectors, of the solutions of the BK equation is shown in Fig.~\ref{fig:theta} for different rapidities. The range of rapidities  roughly corresponds to the region that can be covered by existing experimental results or by those expected in the near and medium term. The figure presents solutions obtained with approach A to non-locality. The other two approaches to non-locality, B and C,   produce similar behaviour. 
At small impact parameters the effects of the asymmetry are less visible than at large impact parameters. This is understandable because the initial condition, see Eq.~\ref{eq:N0}, is relatively homogeneous for small values of $b$, so the orientation of the dipole does not play such a big role. The evolution slowly washes out the asymmetry.

\begin{figure}[!t]
\centering 
\includegraphics[width=0.48\textwidth]{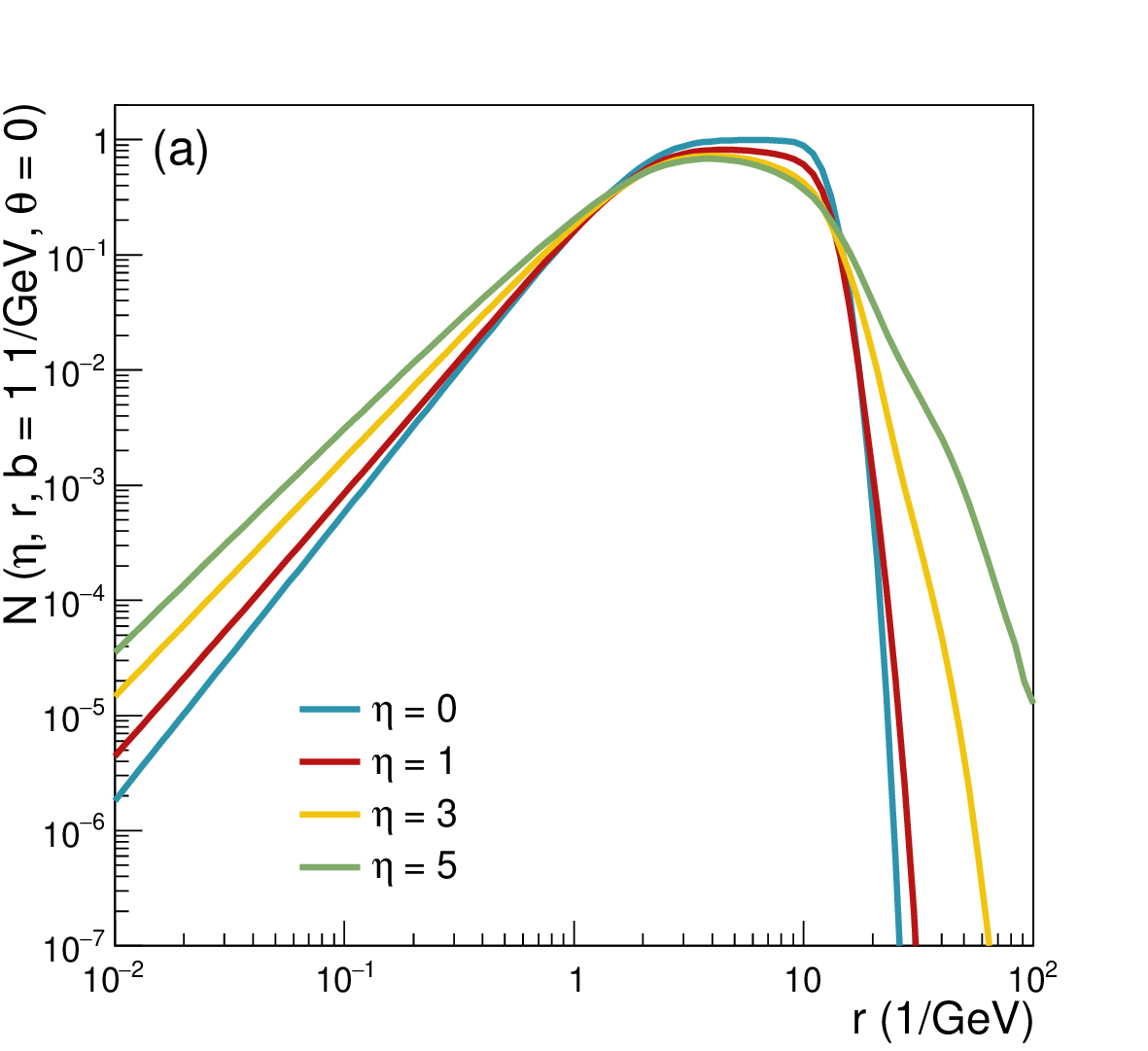}
\includegraphics[width=0.48\textwidth]{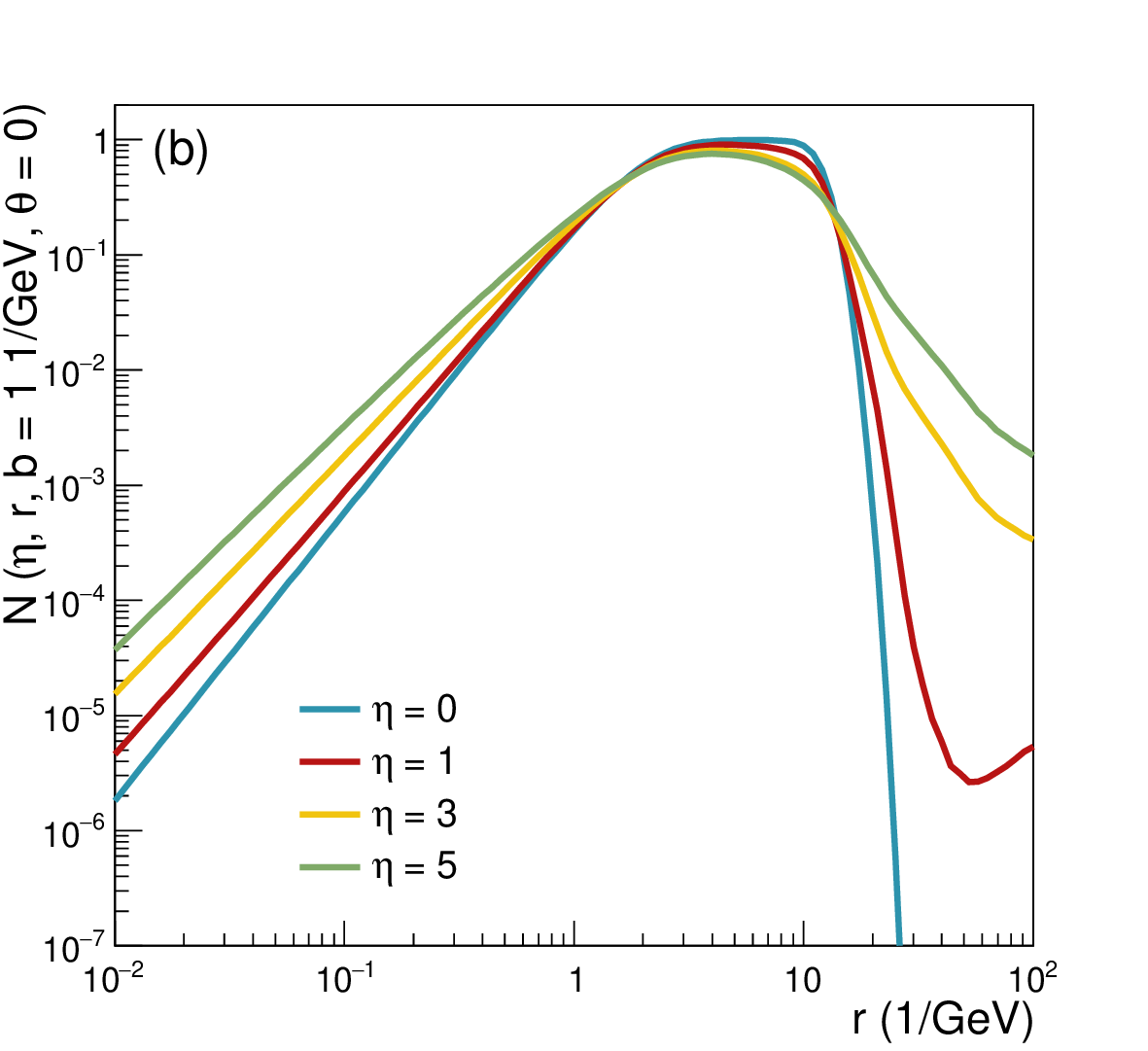}
\caption{Dependence of the dipole amplitude on dipole size $r$  at an impact parameter $b=1$ GeV$^{-1}$ and an  angle between the dipole-size and the impact-parameter  vectors $\theta=0$. Solutions are shown at different rapidities for (a) approach A, and (b) approach  B  to non-locality.
\label{fig:rAB} 
}
\end{figure}

The $r$ dependence of solutions of the BK equations are shown in Fig.~\ref{fig:rAB}. The dipole amplitudes are shown at different rapidities at one impact parameter. Other impact parameter values would give a qualitatively similar picture. The dipole amplitudes are shown for two ways of dealing with the non-locality at early rapidities, namely approaches A and B. In both cases, a wave front towards large values of $r$ develops, in addition to the traditional wave front towards small values of $r$. The evolution of the front at large $r$ has been observed and explained in previous studies of impact-parameter dependent solutions of the BK equation~\cite{Golec-Biernat:2003naj,Berger:2010sh}. The basic idea is that for very large dipoles the ends are outside the target and the interaction of the dipole with the target is strongly suppressed for such configurations. That is, this effect is a direct consequence of having a target that is finite in the impact-parameter plane.
The new observation brought up by the non-locality present in the BK equation evolved in target rapidity is that the shape and size of the large-$r$ wavefront depend on the treatment of the region of rapidities earlier than the initial rapidity of the evolution. Approach B shows a fast evolution of the large-$r$ wavefront and the shape of the front is strikingly different from option A.  The results for approach B are qualitatively similar to those of C, with the latter showing an even stronger effect. The behaviour is qualitatively the same at all angles $\theta$.

\begin{figure}[!t]
\centering 
\includegraphics[width=0.48\textwidth]{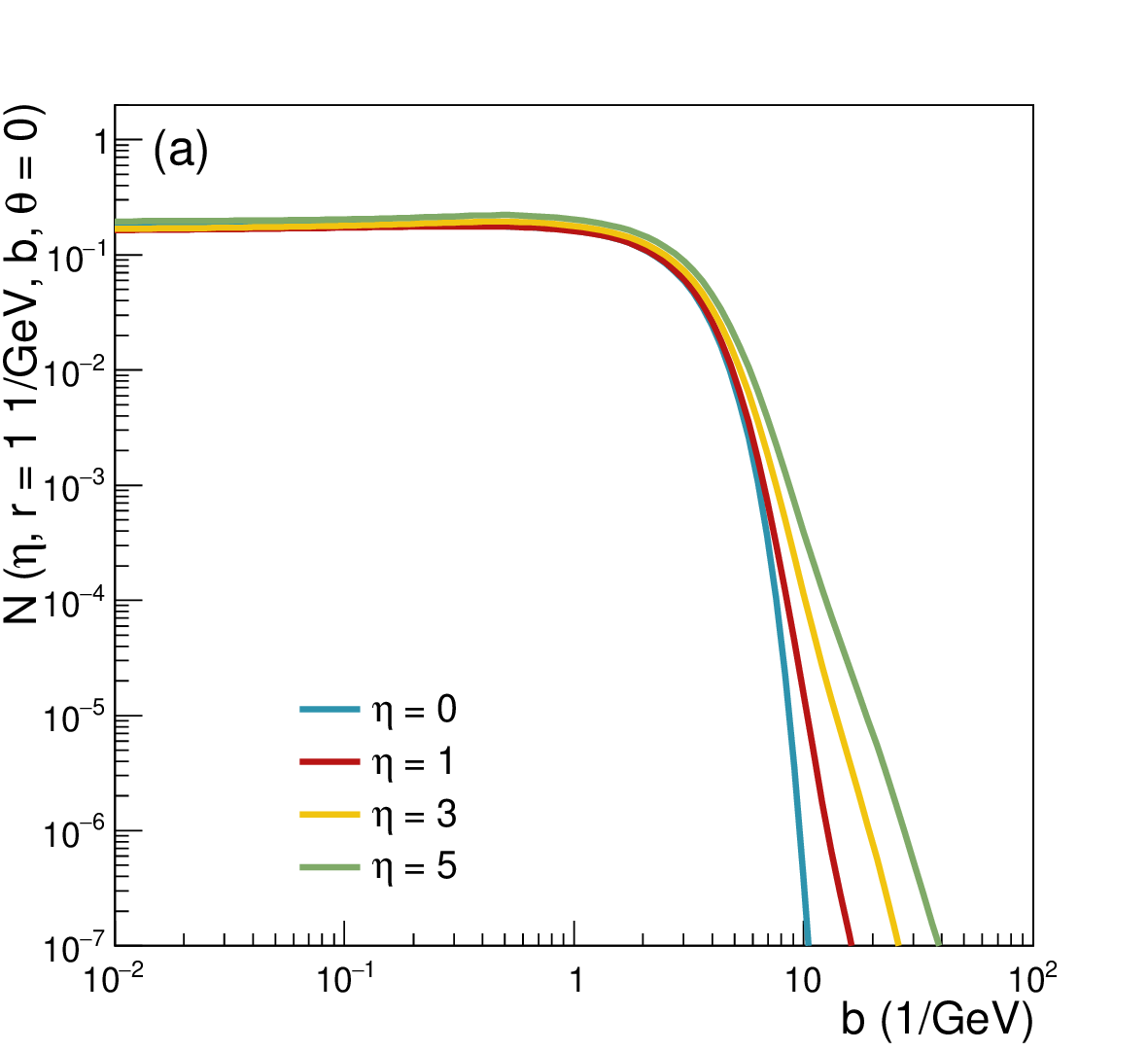}
\includegraphics[width=0.48\textwidth]{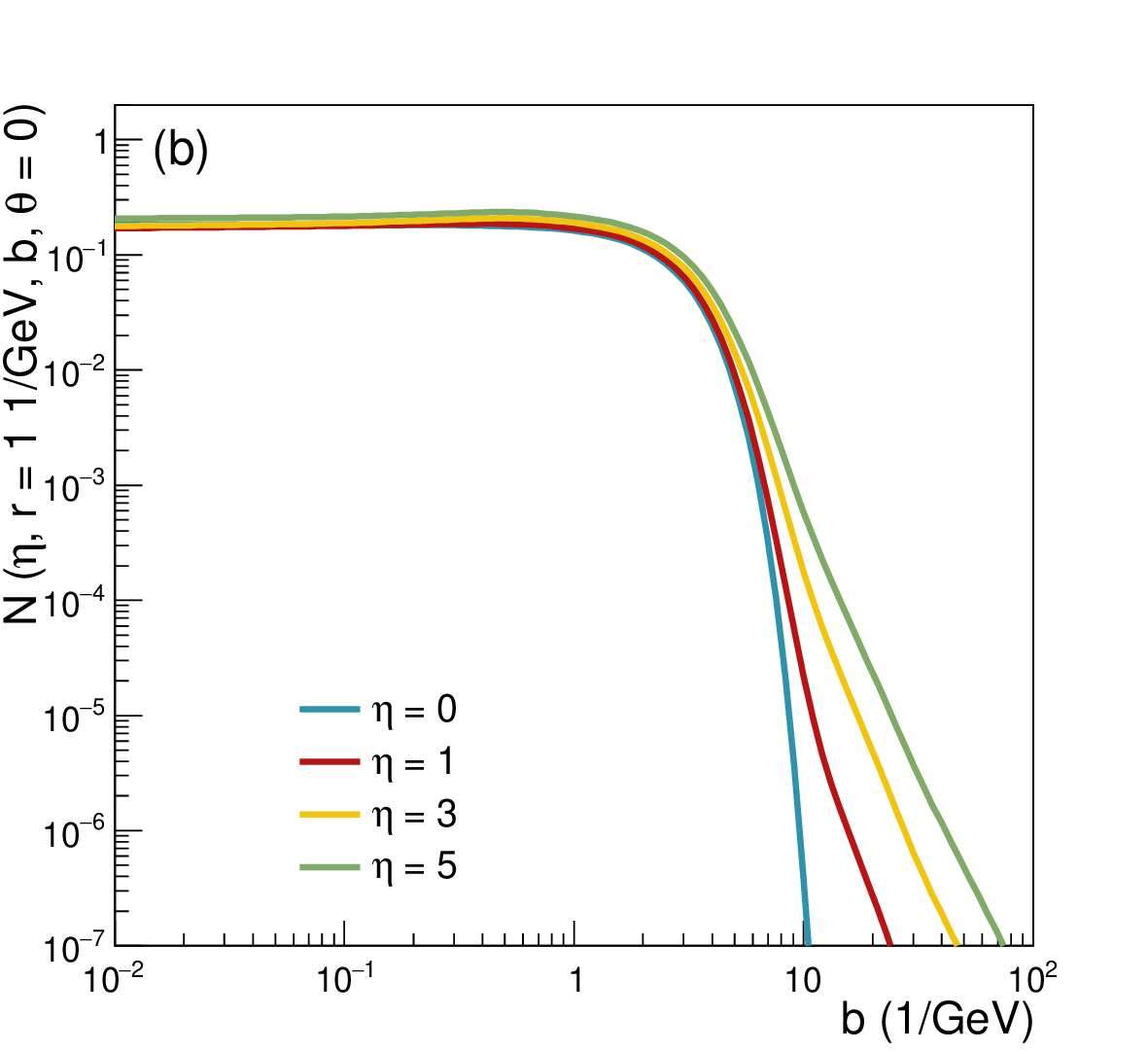}
\caption{Dependence of the dipole amplitude on impact parameter $b$ for a dipole size $r=1$  GeV$^{-1}$ and an angle between the dipole-size and the impact-parameter  vectors $\theta=0$. Solutions are shown at different rapidities for (a) approach A, and (b) approach  B  to non-locality. 
\label{fig:bAB} 
}
\end{figure}

The dependence on impact parameter $b$ of the solutions of the BK equation is shown in Fig.~\ref{fig:bAB} at different rapidities for approaches A and B to non-locality. Approach C produces similar results as B, but with a stronger evolution. The figure shows the case of $\theta =0$. Other angles behave in the same way qualitatively.  The behaviour for other $r$ values is qualitatively as for $r=1$ GeV$^{-1}$ shown in the figure, except for the case of very large dipoles and $\theta=0$ that develop an enhancement at large impact parameters as already observed and explained in the first studies of the impact-parameter dependence of the BK equation~\cite{Golec-Biernat:2003naj}. The figure shows a flat behaviour from small impact parameters up to  $b$ around 5 GeV$^{-1}$, where the dipole amplitude starts to decrease rapidly because the initial condition represents a finite target. The evolution increases the range in impact parameter where the dipole amplitude is sizable, but also it changes the shape of the amplitude at large impact parameters. In particular, it seems that the Coulomb tails are present for approaches B and C to non-locality. Coulomb tails were shown to be  strongly suppressed for $b$-dependent solutions of the BK solutions evolved in the dipole rapidity~\cite{Cepila:2018faq, Bendova:2019psy}. 
The suppression was traced back to the use of the  collinearly improved kernel~\cite{Iancu:2015vea,Iancu:2015joa}, which suppresses large daughter dipoles that live longer than the parent dipole. Evolution in target rapidity is expected not to suffer from these issues, so the kernel has a different form than the collinearly improved kernel. Given that approach A shows suppressed Coulomb tails,  it seems that the tails at large impact parameters observed for approaches B and C are related to the non-locality of the BK equation, and specifically to the influence of what happens at negative rapidities.

\section{Applications to phenomenology
\label{sec:pheno}}

We compared the solutions to the BK equation evolved in target rapidity to HERA measurements of structure functions obtained in deep-inelastic scattering~\cite{H1:2009pze} and the cross section for diffractive exclusive $\Jpsi$ vector meson photo- and electro-production~\cite{H1:2005dtp}. The relation between the dipole amplitude and the observables is given in~\ref{sec:app}. No attempt has been made to fit the parameters to improve the quality of data description. Comparing the value of the parameters shown in Table~\ref{tab:para} with those found in Ref.~\cite{Beuf:2020dxl} one can see that  the values used here are similar to some of the sets of parameters quoted in Ref.~\cite{Beuf:2020dxl}.  

\begin{figure}[!t]
\centering 
\includegraphics[width=0.48\textwidth]{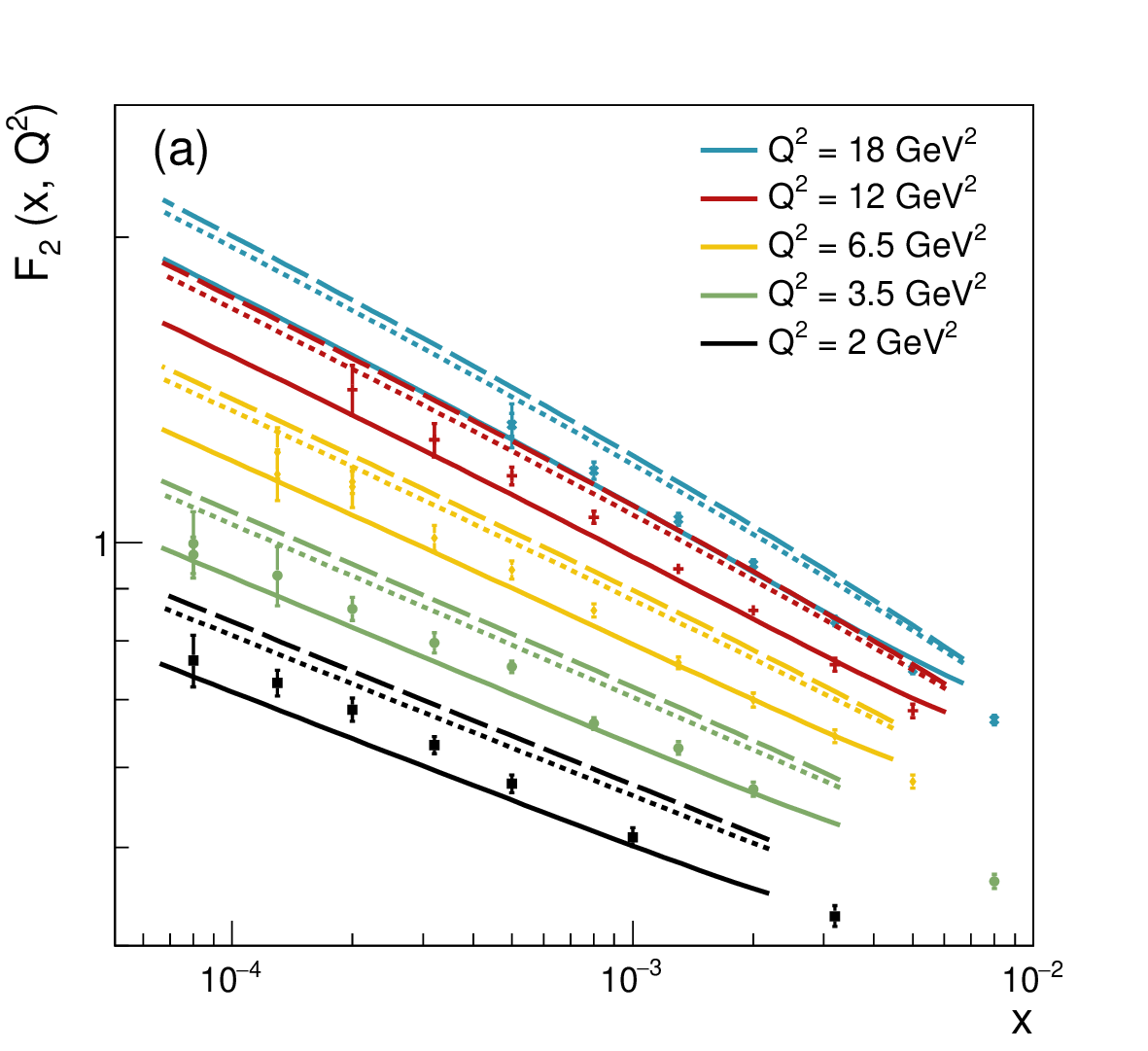}
\includegraphics[width=0.48\textwidth]{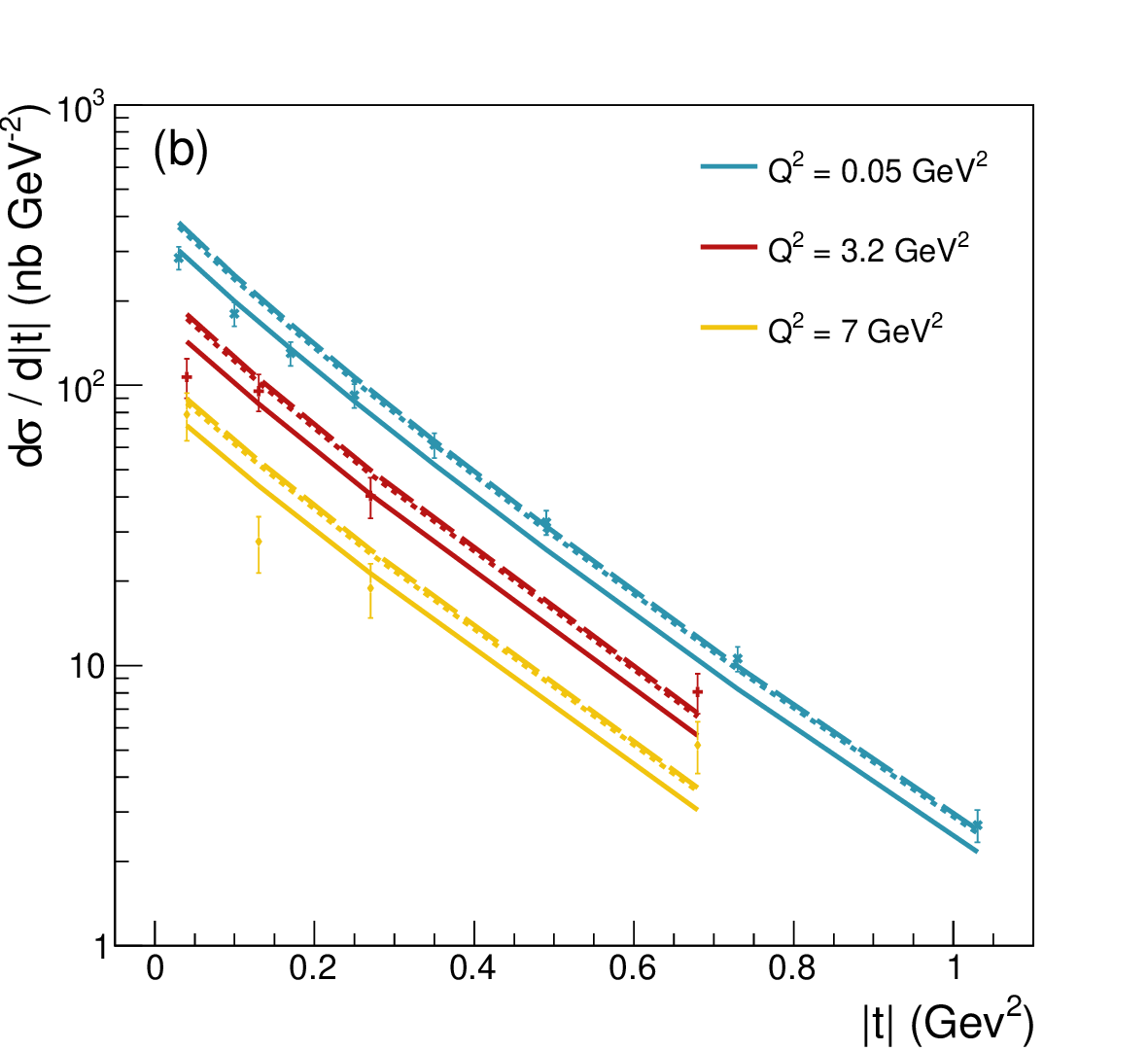}
\caption{Predictions using the BK equations solved with the inclusion of angular correlations between the dipole orientation and the impact parameter and using the target rapidity as the evolution variable. The three approaches to deal with the non-localities for early rapidities are shown with solid (A), dotted (B) and dashed (C) lines. Panel (a) shows the $F_2$ structure function as measured at HERA~\cite{H1:2009pze} and panel (b) shows the cross section for diffractive exclusive $\Jpsi$ vector meson photo- and electroproduction~\cite{H1:2005dtp}.
\label{fig:pheno}}
\end{figure}

The comparison of the predictions of the three approaches to deal with the non-locality of the BK equation in projectile rapidity for early rapidities with data from HERA is shown in Fig.~\ref{fig:pheno}. In all cases shown in the figure, approaches B and C predict a larger cross section than approach A. 
 The  $F_2$ data is reasonably well described by approach A. The same can be said of the comparison to the vector meson data. 
 This demonstrates that also in the case of target evolution, the BK equation can be used for phenomenological applications without adding ad hoc prescriptions to deal with the influence of Coulomb tails, as already found for the case of the impact-parameter dependence of the BK equation evolved in the projectile rapidity~\cite{Cepila:2018faq, Bendova:2019psy}.
   
\section{Summary and outlook
\label{sec:summary}}

We have obtained, for the first time, solutions of the BK equation evolved in the target rapidity including the full impact-parameter dependence. The issue of how to deal with early rapidities was explored using three different approaches. Two of these approaches denoted B and C above, give rise to mild Coulomb tails, while the third approach, denoted A, does not. The solutions have been used to obtain predictions for physical observables, namely the $F_2$ structure function of protons and the cross section for diffractive exclusive photo- and electroproduction of $\Jpsi$ vector mesons off protons.  Both sets of predictions are compared to existing data from HERA and a reasonable agreement is found. For the chosen set of parameters, the by far best agreement is obtained with approach A. 

These results demonstrate that, even though the part of the collinearly improved kernel responsible for the suppression of Coulomb tails is absent,  also for the case of the target-rapidity evolved BK equation it is possible to obtain phenomenologically viable solutions without any extra ad hoc ingredients. This opens the possibility to
use solutions of this equation to explore other observables that are to be measured at current facilities, like RHIC and the LHC, or those that will enter operation in the near future, like the EIC. 

A particularly interesting set of observables to explore the angular dependence between the impact-parameter and the dipole-size vector is that related to the elliptical Wigner distribution as extensively discussed for example in Ref.~\cite{Pasechnik:2023mdd}.

\section*{Acknowledgements}
This work was partially funded by the Czech Science Foundation (GAČR), project No. 22-27262S.

\bibliography{bibliography}

\appendix
\section{Formulas to compute observables \label{sec:app}}

The relation between the dipole amplitude $N$ and the structure function $F_2$ is given by

\begin{equation}
    F_2 (x, Q^2) = \sum\limits_{f} \frac{Q^2}{4\pi^2 \alpha_{\rm em}} \left[ \sigma^{\gamma*p}_{L, f}\left(Q^2, x_f(x, Q^2)\right) + \sigma^{\gamma*p}_{T, f}\left(Q^2, x_f(x, Q^2)\right) \right],
\end{equation}
where $f$ denotes the flavour of a quark, $Q^2$ is the virtuality of the exchanged photon, $ \alpha_{\rm em}$ is the electromagnetic coupling constant, 
\begin{equation}
    x_f = \frac{x_0 e^{-\eta}}{1+4\frac{m_{f}^2}{Q^2}},
\end{equation}
\begin{equation}
    \sigma^{\gamma*p}_{L, T, f}(Q^2, x_f) 
    = 8\pi\int\de{r} r \int\de{z} |\psi_{L, T, f}(r, z, Q^2)|^2 \int \de{b} b\int\limits_{0}^{\pi} \de{\theta} N(r, b, \theta, \eta(x_f)),
\end{equation}
with the longitudinal ($L$) and transverse ($T$) light-cone wave functions
\begin{equation}
  \left| \psi_{L, f}(r, z, Q^2) \right|^2= \frac{N_C \alpha_{em}}{2\pi^2}e_f^2 4 Q^2 z^2 (1-z)^2 K_0^2\left(r \epsilon\right) 
\end{equation}
and
\begin{align}    
    \left|\psi_{T, f}(r, z, Q^2)\right|^2 &= \frac{N_C \alpha_{em}}{2\pi^2}e_f^2 \left[(z^2 + (1-z)^2) (z(1-z)Q^2 + m_{f}^2) K_1^2\left(r \epsilon\right) \right. \nonumber\\
    &\hspace{2.2cm}  \left. + m_{f}^2 K_0^2\left(r \epsilon\right)
    \right]. 
\end{align}
where $\epsilon=\sqrt{z(1-z)Q^2 + m_{f}^2}$, and $K_{0,1}$ are Bessel functions.

The relation between the dipole amplitude and diffractive exclusive vector meson production is given by the sum of the transverse and longitudinal contributions:
\begin{equation}
\frac{\de \sigma_{T,L}}{\de |t|}(t, Q^2, W) =
\frac{1}{16\pi} (1 + \beta_{T,L}^2) R^2_{L,T} \left| \mathcal{A} _{T,L} \right|^2, 
\end{equation}
where 
\begin{equation}
\mathcal{A}(t, Q^2, W) = 
i \int \de \vec{r} \int\limits_{0}^{1} \frac{\de z}{4\pi} \int \de^2 \vec{b} \left( \Psi_V^\dagger \Psi\right)_{T,L}e^{-i[\vec{b}- (\frac{1}{2}-z)\vec{r}]\vec{\Delta}} 2 N\left(r,b, \eta)\right)
\end{equation}
with $x=(Q^2+M^2)/(W^2+Q^2)$, $M$ the mass of the vector meson, $W$ the centre-of-mass energy of the photon--proton system, $\vec{\Delta}^2=-t$, and the wave function of the vector mesons given by the boosted Gaussian model with the values of all parameters as in Ref.~\cite{Bendova:2018bbb}. The masses of the light quarks were taken to be 0.1~GeV$/c^2$ and the mass of the charm quark was taken to be 1.3~GeV$/c^2$. The corrections are 
\begin{equation}
\beta_{T,L} = \tan (\pi \lambda_{T,L}/2), ~~~~~ R_{T,L} = \frac{2^{2\lambda_{T,L}+3}}{\sqrt{\pi}} \frac{\Gamma (\lambda_{T,L} + 5/2)}{\Gamma (\lambda_{T,L} + 4)},
\end{equation}
with 
\begin{equation}
    \lambda_{T,L} = \frac{\partial \ln \mathcal{A}_{T,L}}{\partial \ln (1/x)}.
\end{equation}

\end{document}